# Enhancement of spin transparency by interfacial alloying

Lijun Zhu,[1]* Daniel C. Ralph,[1,2] and Robert A. Buhrman[1]
1. Cornell University, Ithaca, NY 14850
2. Kavli Institute at Cornell, Ithaca, New York 14853, USA

*e-mail: lz442@cornell.edu

We report that atomic-layer alloying (intermixing) at a Pt/Co interface can increase, by approximately 30%, rather than degrade the interfacial spin transparency, and thereby strengthen the efficiency of the dampinglike spin-orbit torque arising from the spin Hall effect in the Pt. At the same time, this interfacial alloying substantially reduces fieldlike spin-orbit torque. Insertion of an ultrathin magnetic alloy layer at heavy-metal/ferromagnet interfaces represents an effective approach for improving interfacial spin transparency that may enhance not only spin-orbit torques but also the spin detection efficiency in inverse spin Hall experiments.

Current-induced spin-orbit torques (SOTs) show promise for driving energy-efficient magnetic memory, nano-oscillators, and non-volatile logic [1-4]. The dominant source of the SOTs in heavy metal/ ferromagnetic layer (HM/FM) systems is usually found to be the spin Hall effect (SHE) of the HM [5-7]. In this case, the dampinglike (DL) efficiency per unit current density ($\xi_{DL}^j$) is the product of the spin Hall ratio ($\theta_{SH}$) within the HM and the spin transparency ($T_{int}$) of the HM/FM interface, *i.e.* $\xi_{DL}^j = T_{int}\theta_{SH}$ [8-19]. It is therefore essential to optimize $T_{int}$ of HM/FM interfaces as well as $\theta_{SH}$ for energy-efficient SOT applications. Two phenomena are generally considered important in limiting $T_{int}$: (i) spin backflow (SBF) that is significant when the bare interfacial spin mixing conductance ($G^{\uparrow\downarrow}$) is comparable to or smaller than the spin conductance ($G_{HM}$) of the HM [8-10], and (ii) interfacial spin memory loss (SML) [11-17] due to spin scattering at the interface. Strong variations in $T_{int}$ are often observed as a function of thermal annealing and/or of deposition conditions, leading to suggestions that these variations are due to enhanced SBF and/or SML by interfacial intermixing [11,12,20]. However, first-principles calculations indicate that the effects of interfacial intermixing on $G^{\uparrow\downarrow}$ should be small, at least in the absence of substantial interfacial spin-orbit coupling (ISOC)(with the prediction being either a small decrease [21] or a modest (~14%) increase [22]), and that intermixing should also not be an important cause of SML [16,23]. Very recent work [17] also show that SML at HM/FM interfaces increases significantly with ISOC in the case of negligible interface intermixing. So far, a direct experimental test as to whether or how interface alloying (intermixing) affects $T_{int}$ and SOT efficiencies has been missing.

In this work, we report a comparative experimental study showing that the effect of interfacial alloying in a Pt/Co heterostructure where the SOTs are due to the SHE of Pt [7,17] is a significant (~30%) enhancement of $\xi_{DL}^j$ due to an improvement of $T_{int}$ of the Pt/FM interface. We attribute this improvement in part to a reduction of ISOC and hence in SML. The rest of the improvement in $T_{int}$ appears to be due to an increase in $G^{\uparrow\downarrow}$.

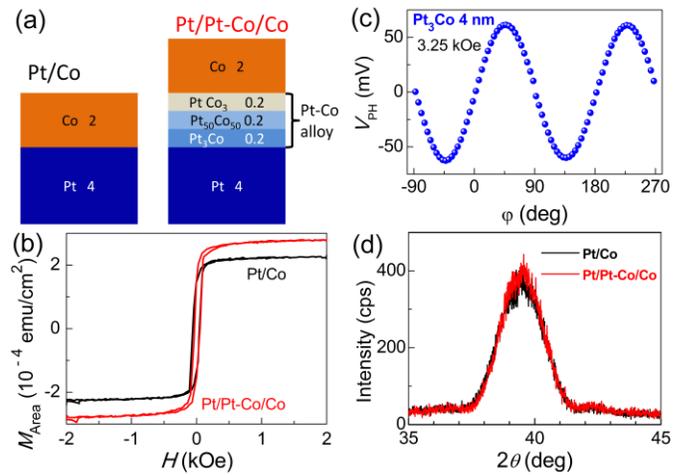

Fig. 1. (a) Schematic of sample layers (not to scale). (b) In-plane magnetization ($M_{area}$) at 300 K versus in-plane magnetic field ($H$). (c) The planar Hall voltage of an isolated 4 nm $Pt_3Co$ single layer plotted as a function of in-plane bias field orientation (field magnitude = 3.25 kOe), showing that the $Pt_3Co$ is magnetic at room temperature. (d) X-ray diffraction peaks for the Pt/Co samples with (red) and without (black) interface alloying (Pt-Co).

For this study we deposited in-plane magnetized stacks of Pt 4/Co 2 bilayers with and without a Pt-Co alloy interface layer consisting of $Pt_3Co$ 0.2/PtCo 0.2/$PtCo_3$ 0.2 (the numbers are layer thicknesses in nm, see Fig. 1(a)) by DC/RF sputtering onto oxidized Si substrates. The trilayer Pt-Co alloy spacer is used to simulate a gradual intermixing at the Pt/Co interface. Each stack is seeded by a 1 nm Ta layer a to improve the adhesion and smoothness of the subsequent Pt and Co layers, and capped by a 2 nm MgO layer and finally a 1.5 nm Ta layer that was fully oxidized upon exposure to atmosphere. No thermal annealing was performed on the samples. We measured the in-plane magnetic moment per unit area ($M_{area}$) of each sample at 300 K by sweeping the magnetic field ($H$) up to 3.5 T along film plane using a vibrating sample magnetometer. Figure 1(b) shows the in-plane $M_{area}$-$H$ curves in the low field range (< 2 kOe), from which one can see that both the Pt/Co and the Pt/Pt-Co/Co samples exhibit a small region of hysteresis near zero field followed by a quick saturation of magnetization, indicating in-plane magnetic anisotropy. Using the saturated value of $M_{area}$ for the Pt/Co bilayer sample, we determine the average saturation magnetization per unit volume ($M_s$) for the Co layer to be $1192 \pm 10$ emu/cm$^3$ (~1.43 $\mu_B$/Co), lower than 1440 emu/cm$^3$ (~1.74 $\mu_B$/Co) for bulk Co [24]. Compared to the Pt/Co bilayer sample, the saturated value of $M_{area}$ for the Pt/Pt-Co/Co sample is enhanced by ~$0.53\times10^{-4}$ emu/cm$^2$, indicating an *average* $M_s$ of $915 \pm 29$ emu/cm$^3$ ($2.19\pm0.07$ $\mu_B$/Co) for the Pt-Co alloy spacer layer, which is suggestive of an enhanced magnetic proximity effect due to contact of Co with Pt [18] (*e.g.*, ~2.13 $\mu_B$/Co for $L1_0$-Pt$_{50}$Co$_{50}$ [25]). We also find that even the most dilute alloy (Pt$_3$Co) is still magnetic at room temperature as indicated by the planar Hall signals in a 4 nm thick Pt$_3$Co control sample (Fig. 1(c)). The interface alloying has no discernible influence on the strain of the (111)-textured Pt layers as indicated by the good overlap of the Pt x-ray diffraction (XRD) patterns for the stacks within and without the Pt-Co alloy spacer (Fig. 1(d)).

The samples were further patterned into $5\times60$ μm$^2$ Hall bars for determination of the DL and the fieldlike (FL) SOT efficiencies by harmonic Hall response measurements under a sinusoidal electric bias field of $E = 66.7$ kV/m (more details of measurement geometry can be found in our previous papers [7,17]). For in-plane magnetized HM/FM bilayers [7,17], the dependence of out-of-phase second harmonic Hall voltage ($V_{2\omega}$) on the in-plane field angle ($\varphi$) is given by

$$V_{2\omega} = V_a\cos\varphi + V_p\cos\varphi\cos2\varphi, \quad (1)$$

where $V_a = -V_{AH}H_{DL}/2(H_{in}+H_k) + V_{ANE}$, and $V_p = -V_{PH}(H_{FL}+H_{Oe})/H_{in}$, with $H_{DL(FL)}$ being DL(FL) SOT fields, $V_{AH}$ the anomalous Hall voltage, $V_{ANE}$ the anomalous Nernst voltage, $H_{in}$ the in-plane bias field, $H_k$ the perpendicular anisotropy field, $V_{PH}$ the planar Hall voltage, and $H_{Oe}$ the Oersted field. We separated the DL term $V_a$ and the FL term $V_p$ for each value of $H_{in}$ by fitting the $V_{2\omega}$ data to Eq. (1) (see Fig. 2(a)). The linear fits of $V_a$ versus $-V_{AH}/2(H_{in}+H_k)$ and $V_p$ versus $-V_{PH}/H_{in}$ (see Fig. 2(b)) yield the values of $H_{DL}$ and $H_{FL}$. We note that the thermal effects [26] (i.e. the ordinary Nernst effect in the Pt and Co layers and the anomalous Nernst effect in Co) are negligible in the conductive Pt/Co and Pt/Pt-Co/Co samples as evidenced by the good linearity and the small intercepts (i.e. $V_{ANE}$) in the fits of $V_a$ versus $-V_{AH}H_{DL}/2(H_{in}+H_k)$. Using the values of $H_{DL(FL)}$, the DL(FL) SOT efficiencies per applied electric field can be determined as

$$\xi^E_{DL(FL)} = (2e/\hbar)\mu_0 M_{area}H_{DL(FL)}/E, \quad (2)$$

with $e$, $\hbar$ and, $\mu_0$ being the elementary charge, the reduced Planck constant, and the permeability of vacuum. Correspondingly, the SOT efficiencies per unit bias current density are

$$\xi^j_{DL(FL)} = (2e/\hbar)\mu_0 M_{area}H_{DL(FL)}/j_e, \quad (3)$$

where $j_e = E/\rho_{xx}$ is the charge current density and $\rho_{xx}$ is the longitudinal resistivity of the spin Hall channel (*i.e.*, Pt here). $\rho_{xx}$ for the Pt layer was determined to be ≈ 42.5 μΩ cm for both samples with and without the Pt-Co interfacial layer. The final results for both $\xi^E_{DL(FL)}$ and $\xi^j_{DL(FL)}$ for the Pt/Co and Pt/Pt-Co/Co samples are shown in Figs. 2(c) and 2(d).

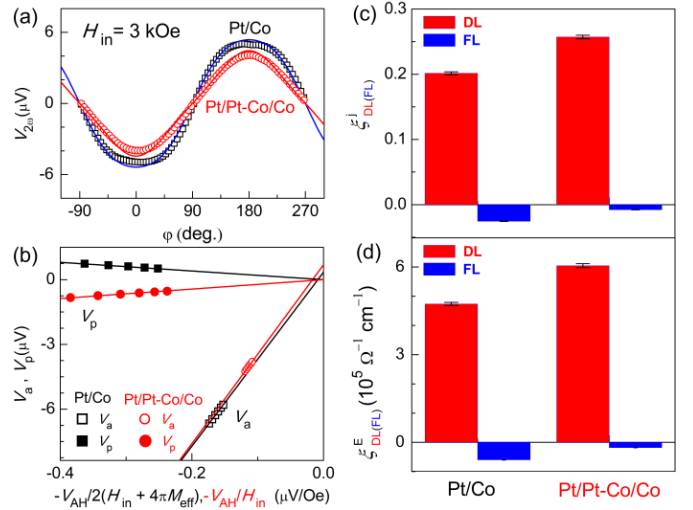

Fig. 2. (a) $\varphi$ dependence of $V_{2\omega}$; (b) $V_a$ versus $-V_{AH}/2(H_{in}+H_k)$ and $V_p$ versus $-V_{PH}/H_{in}$; (c) the SOT efficiencies per unit bias current density $\xi^j_{DL(FL)}$ and (d) the SOT efficiencies per unit applied electric field $\xi^E_{DL(FL)}$ for the Pt/Co and Pt/Pt-Co/Co samples.

We first focus on the DL SOT efficiencies, the quantity of primary importance for efficiently driving magnetization switching and chiral domain wall displacement. Rather than finding a degradation arising from an intermixed interface, we find that, $\xi^E_{DL}$ is enhanced by more than 28%, from $(4.74\pm0.05)\times10^5$ $\Omega^{-1}$ m$^{-1}$ for Pt/Co to $(6.05\pm0.07)\times10^5$ $\Omega^{-1}$ m$^{-1}$ for Pt/Pt-Co/Co; correspondingly, $\xi^j_{DL}$ is enhanced from $0.201\pm0.002$ for Pt/Co to $0.257\pm0.003$ for Pt/Pt-Co/Co. The large $\xi^j_{DL}$ of ≈ 0.2 for the Pt/Co bilayer is consistent with that reported for resistive Pt [7,17-19], while significantly greater than the values previously reported for lower-resistivity Pt (e.g., $\xi^j_{DL}$ =0.08 for Ni$_{81}$Fe$_{19}$ 6/Pt 6 with $\rho_{xx}$ = 20 μΩ cm [27]; $\xi^j_{DL}$ =0.12-0.15 for Ta 2/Pt 4/Co wtih $\rho_{xx}$ = 28 μΩ cm [12]). This can be understood as due to an enhancement of $\theta_{SH}$ for the intrinsic SHE with increasing $\rho_{xx}$ [6,7,19] along with reduced SBF due to a reduced spin diffusion length ($\lambda_s$)[28,29]. Nguyen *et al.* [6] reported a $\xi^j_{DL}$ of ≤ 0.12 for a highly resistive Pt 4/Co bilayer ($\rho_{xx}$=50 μΩ cm); however, in those samples ISOC and thus SML [17] were particularly strong.

Compared with the Pt/Co sample, the insertion of the interfacial alloy layer in the Pt/Pt-Co/Co sample has no discernible influence on the crystalline structure or $\rho_{xx}$ of the Pt layer, indicating that both $\theta_{SH}$ and $\lambda_s$ in the Pt should be the same for both samples. Moreover, the ultrathin Pt-Co alloy layer, which is magnetic and ferromagnetically coupled to the adjacent Co layer (see Figs. 1(b) and 1(c)), should act as an absorber for the spin current generated by the Pt layer. Even if the ultrathin magnetic Pt$_3$Co 0.2/PtCo 0.2/PtCo$_3$ 0.2 layers were treated as a nonmagnetic spin current generator that is as efficient as an additional 0.6 nm Pt, the effect is still too small to explain the 28% increase of $\xi^E_{DL}$. As indicated by a thickness-dependence study [6], $\xi^E_{DL}$ for a Pt 4.6/Co bilayer is at most 5% larger than that for a Pt 4/Co bilayer. We also measured the harmonic Hall response for a control sample with the structure Pt$_3$Co 4 nm/Co 2 nm. The effective DL harmonic Hall signal was at least factor of 6 smaller and of opposite sign relative to the Pt/Pt-Co/Co samples, reaffirming that the Pt-Co alloy layer is not itself a source of the improved SOT in the Pt/Pt-Co/Co sample. Therefore, the enhancement of the DL SOT should be attributed to an improved $T_{int}$, i.e. the deliberate introduction of interface disorder reduces the ISOC and/or increases $G_{eff}^{\downarrow\uparrow}$ of the Pt/Co interface.

To quantify the possible reduction in SML, we first determined the strength of the interfacial magnetic anisotropy energy density ($K_s$), a linear indicator of the strength of ISOC [17]. We calculated the magnetic anisotropy field ($4\pi M_{eff}$) from ferromagnetic resonance measurements and Kittel's formula [30],

$$f = (\gamma/2\pi)\sqrt{H_r(H_r + 4\pi M_{eff})} \quad (4)$$

where $f$ is the rf frequency, $\gamma$ the gyromagnetic ratio, and $H_r$ the ferromagnetic resonance field. As shown in Fig. 3(a), $4\pi M_{eff}$ increases from 0.375 ± 0.004 T for the Pt/Co to 0.682 ± 0.002 T for Pt/Pt-Co/Co. Using the relation $4\pi M_{eff} \approx 4\pi M_s - 2K_s/M_{area}$, where $M_{area}$ includes the contributions from both the Co and Pt-Co alloy layers, we determine $K_s$ to be 1.335 ± 0.004 erg/cm$^2$ for the Pt/Co sample and 1.062 ± 0.003 erg/cm$^2$ for Pt/Pt-Co/Co sample. After subtracting the contribution of 0.56 erg/cm$^2$ from Co/MgO interface as measured previously [17], we estimate that $K_s$ for the Pt/FM interface ($K_s^{ISOC}$) is reduced from 0.78 erg/cm$^2$ for the Pt/Co to 0.50 erg/cm$^2$ for Pt/Pt-Co/Co interface, indicative of a corresponding decrease in the ISOC strength [31] with the interfacial alloying. This observation is consistent with previous experiments that alloying at the Pt/Co interface reduces its PMA [32,33].

Our recent finding [17] has established a linear degradation of the DL torque with increasing ISOC strength at HM/FM interfaces. Specifically, $\xi_{DL}^E = 5.85 - 1.34 K_s^{ISOC}$ for in-plane magnetized Pt/Co bilayers, with $\xi_{DL}^E$ in the unit of $10^5$ $\Omega^{-1}$ m$^{-1}$ and $K_s^{ISOC}$ in the unit of erg/cm$^2$. For the Pt/Co sample the values of $K_s$ and $\xi_{DL}^E$ are in good agreement with this linear dependence [17], but the value $\xi_{DL}^E = 6.05 \times 10^5$ $\Omega^{-1}$ m$^{-1}$ for the Pt/Pt-Co/Co sample is considerably above the value $5.18 \times 10^5$ $\Omega^{-1}$ m$^{-1}$ predicted by this formula for $K_s^{ISOC}$ = 0.50 erg/cm$^2$, and even larger than the extrapolated result for a Pt/Co interface with zero ISOC ($5.85 \times 10^5$ $\Omega^{-1}$ m$^{-1}$) by 15% even though the interfacial $K_s$ in this case is well above zero (~ 0.50 erg/cm$^2$). Therefore, there should be an additional factor responsible for part of the increase in $\xi_{DL}^E$ between the Pt/Co and alloyed Pt/Pt-Co/Co interface result, with this most likely being an enhancement of Re$G^{\downarrow\uparrow}$.

Although initial theoretical studies indicated that interfacial disorder should have only minimal effect on the real part of spin mixing conductance Re$G^{\downarrow\uparrow}$, e.g. < 5% for a Cu/Co interface [21], a more recent first-principles calculation for Pt (111)/Ni$_{81}$Fe$_{19}$ interface [22] has indicated that Re$G^{\downarrow\uparrow}$ may be increased by as much as 14% by the introduction of two disordered atomic layers. Taken together with our calculation of the reduction of SML due to lower ISOC in the sample with interfacial alloy this effect could account for the overall enhancement (by 30%) in interfacial transparency. Another phenomenon that could in principle affect our results is spin fluctuations – recent experiments have indicated that spin fluctuations due to the insertion of an antiferromagnetic CoO layer at Pt/Co or Pt/YIG interfaces can enhance Re$G^{\downarrow\uparrow}$ [34-36]. If the Curie temperature of the thin layer of Pt$_3$Co in our samples is sufficiently low to allow spin fluctuations, this might also increase Re$G^{\downarrow\uparrow}$. However, spin fluctuations are expected [37] to also increase magnetic damping whereas we do not observe such an increase upon introduction of the interfacial alloy layers. This indicates that spin fluctuations, if any, should be a minor effect here.

While generally less important technologically the FL SOT is certainly of fundamental interest. For cases where the FL SOT arises from an incident spin current due to the bulk spin Hall effect, it is understood to be the result of spin rotation upon reflection from the interface region and to be linearly proportional to the imaginary part of the interfacial spin mixing conductance (Im$G^{\downarrow\uparrow}$). For the Pt/Co interface it has been calculated [9,10] that $|\xi_{FL}^{E(j)}/\xi_{DL}^{E(j)}|=|$Im$G^{\downarrow\uparrow}$/Re$G^{\downarrow\uparrow}|\approx$ 0.15 [9,10]. In the absence of the interface alloying, we find that $|\xi_{FL}^{E(j)}/\xi_{DL}^{E(j)}|\approx 0.13$, quite consistent with Pt/Co prediction and with the FL torque in Pt/Co system being dominated by the spin current generated by the bulk SHE of the Pt layer [5,17]. After insertion of the Pt-Co alloy layer at the interface, the FL SOT decreases by a factor of ~3 in magnitude while the DL torque increases by 28%. This represents an interesting suppression of the interfacial spin rotation scattering. The change of the FL torque here cannot be explained as due simply to variation of the ISOC, because in that case the FL torque should scale similarly to the DL torque [17]. Instead, we attribute the reduction of the FL torque to a variation of Im$G^{\downarrow\uparrow}$ for the Pt/Co interface due to the alloying.

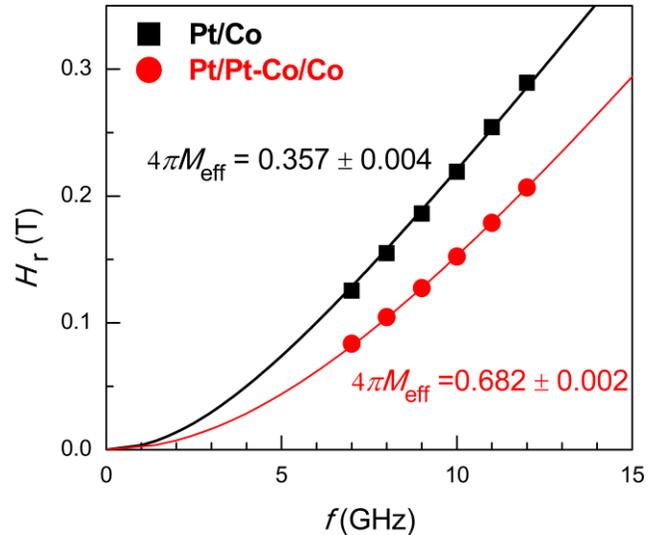

Fig. 3. Influence of interface alloying on the frequency ($f$) dependence of the resonance fields ($H_r$) for the Pt/Co and the Pt/Pt-Co/Co samples. The solid lines in represent the best fits of the data to Eq. (4).

In summary, we have demonstrated that interface alloying in a spin Hall channel/FM interface can enhance rather than degrade the interfacial spin transparency and the DL torque. We find that alloying of the Pt/Co interface reduces SML as indicated by the reduction of the ISOC strength. At the same time, the interface alloy layer appears to moderately enhance Re$G^{\downarrow\uparrow}$ at the interface. We also find that the interface alloying results in a reduction in the FL torque, suggesting an interesting change in Im$G^{\downarrow\uparrow}$ with alloying of the Pt/Co interface. Our results advance the understanding of the effect of interface alloying (intermixing) on spin transport across a HM/FM interface. This work also indicate that insertion of an interface alloy layer can be an effective approach to enhance the spin transparency of a spin Hall channel/FM interface and to optimize current-induced SOTs and spin detection efficiency in an inverse spin Hall experiments.

This work was supported in part by the Office of Naval Research (N00014-15-1-2449), in part by the NSF MRSEC program (DMR-1719875) through the Cornell Center for Materials Research, and in part by the NSF (ECCS-1542081) through use of the Cornell Nanofabrication Facility/National Nanotechnology Coordinated Infrastructure.